\RequirePackage{fix-cm}
\documentclass[smallextended]{svjour3} 
\usepackage{subfig}
\usepackage{graphicx}
\usepackage{float}
\usepackage{url}
\graphicspath{{2/}}
\usepackage{mathptmx}
\usepackage[T1]{fontenc}
\usepackage{times}
\usepackage{xcolor}
\usepackage{algorithmic}
\usepackage{algorithm}
\usepackage{footnote}
\usepackage{threeparttable}
\usepackage{multirow}
\usepackage{chngcntr}
\usepackage[cmex10]{amsmath}
\usepackage[justification=centering]{caption}

\usepackage{blkarray}
\usepackage{amssymb}
\usepackage{todonotes}
\usepackage{aps-bibstyle} 
\title{Link Prediction Approach to Recommender Systems}
\author{T. Jaya Lakshmi         \and
       S. Durga Bhavani 
}
\institute{T. Jaya Lakshmi \at
             School of Engineering and Applied Sciences \\
             SRM University - Andhra Pradesh, India.\\
           \email{jaya.phd.hcu@gmail.com}   
      \and
     S. Durga Bhavani \at
    School of Computer and Information Sciences \\
      University of Hyderabad\\
     Hyderabad, India\\
    \email{sdbcs@uohyd.ernet.in}
}
\begin{document}
\maketitle	
\begin{abstract}
The problem of recommender system is very popular with myriad available solutions. A novel approach that uses the link prediction  problem in social networks has been proposed in the literature that model the typical user-item information as a bipartite network in which link prediction would actually mean recommending an item to a user. The standard recommender system methods suffer from the problems of sparsity  and  scalability. Since link prediction measures involve computations pertaining to small neighborhoods in the network, this approach would lead to a  scalable solution to recommendation. One of the issues in this conversion is that link prediction problem is modelled as a binary classification task whereas the problem of recommender systems is solved as a  regression task in which the rating of the link is to be predicted. We overcome this issue by predicting top k links as recommendations with high ratings without predicting the actual rating. Our work extends similar approaches in the literature by focusing on exploiting the probabilistic measures for link prediction.   Moreover,  in the proposed approach,  prediction measures that utilize temporal information available on the links prove to be more effective in improving the accuracy of prediction.  This approach is evaluated on the benchmark 'Movielens' dataset. We show that the usage of temporal probabilistic measures helps in improving the quality of recommendations.  Temporal random-walk based measure T\_Flow improves recommendation accuracy by 4\% and Temporal cooccurrence probability measure improves prediction accuracy by 10\% over item-based collaborative filtering method in terms of AUROC score.
\end{abstract}

\section{Introduction}

Many e-commerce websites provide a wide range of products to the users. The users commonly have different needs and tastes based on which they buy the products. Providing the most appropriate products to the users make the buying process efficient and improves the user satisfaction. The enhanced user satisfaction keeps the user loyal to the website and improves the sales and thus profits to the retailers. Therefore, more retailers start recommending the products to the users which needs  efficient analysis of user interests in products. E-commerce leaders like Amazon and Netflix use recommender systems to recommend products to the users. The standard recommendation algorithms like content-based filtering and collaborative filtering, model the ratings given by users to items as a matrix and predict the ratings of the customers to unrated items based on user/item similarity. 

Recommender systems recommend items to users. Items include products and services such as movies, music, books, web pages, news, jokes and restaurants. The recommendation process utilizes data including user demographics, product descriptions and the previous actions of users on items like buying, rating, and watching.  The information can be acquired explicitly by collecting ratings given by users on items or implicitly by monitoring user's behavior such as songs listened in music websites, news/movies watched in news/movies websites, items bought in e-commerce websites or books read in book-listing websites in the past. Usage of intelligent recommendation systems improved the revenue of Amazon by 35\%, caused a business growth of 24\% for BestBuy, increased 75\% of views on Netflix and 60\% views on YouTube. \cite{chui2017artificial}. Therefore, building a personalized recommendation system has a profound significance not only in commercial arena, but also in the fields like health care, news, food, academia and so many. Each domain needs to consider different features. 

Recommender system is a typical application of link prediction in heterogeneous networks. Link Prediction problem infers future links in a network represented as graph where nodes are users and edges denote interactions between users. In the context of recommender systems, the nodes may be of two types: items and users. A transaction of a user buying an item can be shown as an edge from user node to item node. Recommendation problem can be viewed as a task of selecting unobserved links for each user node, and thus can be modeled as a link prediction problem.

In this work, we have applied various link prediction measures on the bipartite network in the context of recommender systems and verified the efficacy of those measures. We have chosen a medium sized dataset of MovieLens1M for experimentation.
The contributions made in this paper are
	\begin{itemize}
		\item Evaluated the efficacy of base line link prediction measures to the recommendation task.
		\item Extended existing probabilistic measures called co-occurrence probability measure and temporal co-occurrence probability measure to make it suitable for recommendation problem.
	\end{itemize}

\section{Problem Statement}
Schafer et al \cite{RS_base_2001} define the problem of recommender systems as follows:

Given a set of users $U=\lbrace u_1,u_2,\ldots u_m \rbrace $, and items $I= \lbrace I_1, I_2, \ldots ,I_n \rbrace $ and the ratings $R$ representing the ratings given by user $u_i$ to item $I_j$, the task of recommender systems is to recommend a new item $I_j$ to a user $u_i$ which the user $u_i$ did not buy.

For example, consider the matrix given in Fig.\ref{ratings_matrix}, where rows correspond to the users and the columns denote products. The matrix entry $R_{ij}$ represents the rankings given by user $u_i$ on item $I_j$. The main task of the recommender systems is to predict the unrated entries in the rating matrix. 

\begin{figure}
	\begin{equation*}
		\mathbf{R}=
		\begin{blockarray}{*{4}{c} l}
			\begin{block}{*{4}{>{$\footnotesize}c<{$}} l}
				$item_1$ & $item_2$ & $item_3$ & $item_4$ & \\
			\end{block}
			\begin{block}{[*{4}{c}]>{$\footnotesize}l<{$}}
				5 &   &  2 &    & User 1 \\     
				4 & 4 &  2 &    & User 2 \\      
				4  &   &   &   1 & User 3 \\
				& 2 &   &  2   & User 4 \\
			\end{block}
		\end{blockarray}  
	\end{equation*}
	\caption{Example rating matrix denoting rating given by users to 4 products.}
	\label{ratings_matrix}
\end{figure}

Recommender systems are natural examples of weighted bipartite networks. A \textbf{Bipartite network} contains exactly two types of nodes and single type of edges existing between different types of nodes. Bipartite network is defined as  $G = (V_1 \cup V_2, \quad E)$, where $V_1$ and $V_2$ are sets of two types of nodes,  $E$ represents the set of edges between nodes of type $V_1$ and $V_2$. Fig.\ref{user-item-net} depicts a sample scenario of users buying items in e-commerce sites such as Amazon, modeled as a bipartite network. 

\begin{figure}
	\label{user-item-net}
	\centering
	\includegraphics[width=0.4\textwidth]{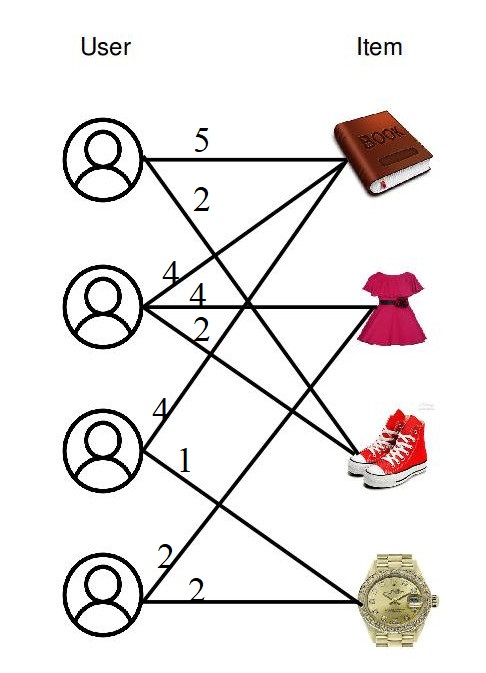}    
	\caption{\small{An example of user-item Bipartite Network with edges denoting rating given by user on item.}} \label{user-item-net}   
\end{figure}

\section{Related Literature}
\begin{figure}
	\centering
	\includegraphics[width=1\textwidth]{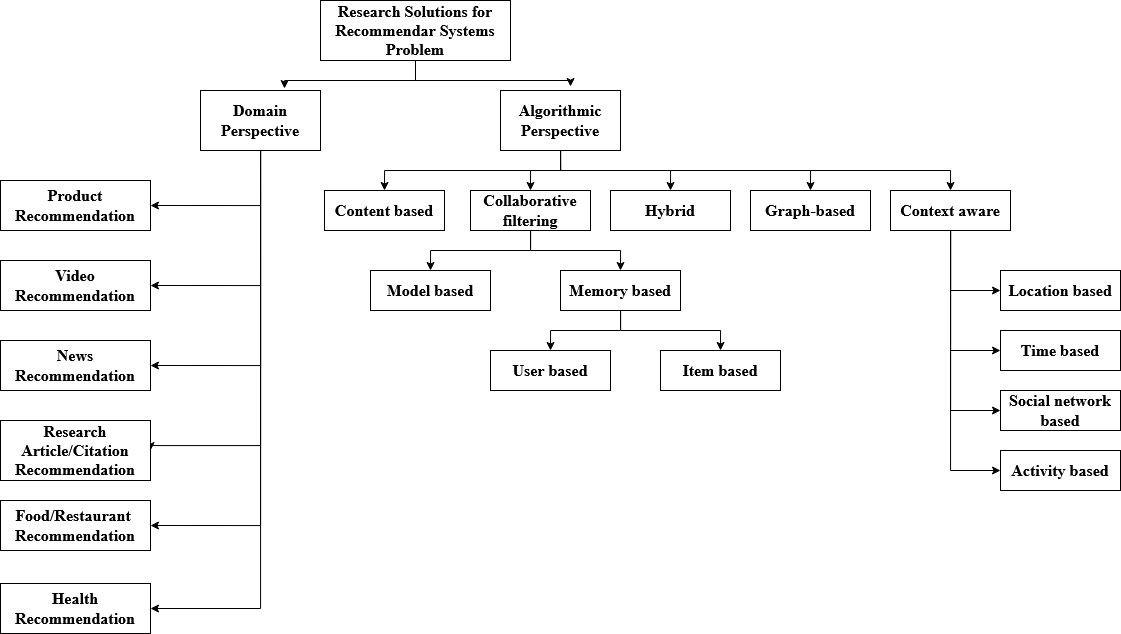}
	\caption{Taxonomy of recommender systems}
	\label{rs_taxonomy}
\end{figure}

The papers on recommender system in the literature are discussed from algorithmic as well as domain perspective in the next two sections. Fig. \ref{rs_taxonomy} gives a taxonomy of the literature.

\subsection{Domain Perspective}
Product recommendation is the mostly explored domain. E-commerce sites like Amazon, Flipkart, Ebay etc use various collaborative filtering techniques for recommending products to their customers. Video recommendations like movies, TV shows and web series are increasing exponentially by companies like Netflix, YouTube. Netflix has launched a competition with 1Million dollars prize money for improving 10\% of RMSE in the movie recommendation, by releasing a 100 million customer ratings~\cite{bell2007lessons}. A news recommendation system has been implemented in~\cite{liu2010personalized} using google news, by constructing a Bayesian network for identifying the user interests and then applying content based as well as collaborative filtering and hybrid techniques. Research article recommendation, food recommendation and health recommendation are other domains among many. 

Valdez et al. integrate recommender systems into individual medical records to help patients  improve their autonomy. The authors have collected health related information from the patients' records and have performed text processing for extracting features and evaluated using metrics available in Information Extraction domain~\cite{valdez2016recommender}.

All these domains need different pre-processing techniques, algorithms, and evaluation metrics. 

\subsection{Algorithmic perspective}
Recommender systems is seeing an explosive growth in the literature with many novel algorithms being proposed almost every day. Content-based recommendation, Collaborative Filtering (CF) and Hybrid approach are some of the popular approaches that  solve the problem of recommender systems. Context aware recommender systems predict recommendations based on user context. Graph based recommender systems is main focus of this work. In graph based recommender system, the purchases are modeled as bipartite graph and various graph traversal techniques are applied to generate a set of co-purchased items/users.  

\subsubsection{Content-based recommendation} 
Content-based recommendation uses item descriptions and constructs user profiles which contain the information about user preferences~\cite{RS_Content_2007}. These user preferences may include a genre, director and actor for a movie, an author for book etc. The recommendation of an item to the user is then based on the similarity between the item description and the user profile.  This method has the benefit of recommending  items to users that have never been rated  by them~\cite{RS_Content_2000}. Content-based recommendation systems require complete descriptions of items and detailed user profiles. This is one of the main limitations of such systems~\cite{RS_Content_2000}. 

Content-based recommender systems are divided into three general approaches: Explicit, Implicit and Model-based. In explicit content based methods,  profile information of the users is collected through questionnaires that ask questions about their preferences for the items. Implicit methods build the user profiles implicitly by searching for similarities in liked and disliked item descriptions by the user. The third class of methods, viz., model-based methods construct user profiles based on a learning method, using item descriptions as input features to be given to  a supervised learning algorithm, and producing user ratings of items as output.

The item descriptions are often textual information contained in documents, web sites and news messages. User profiles are modeled as weighted vectors of these item descriptions. The advantage of this method is the ability to  recommend newly introduced items to users~\cite{RS_Content_2000}. Content-based recommendation systems require the complete descriptions of items and detailed user profiles. This is the main limitation of such systems. Privacy issues such as users dislike to share their preferences with others is another limitation of content-based systems.

\subsubsection{Collaborative filtering}
Collaborative filtering systems, however, do not require such difficult to come by, well-structured item descriptions. Instead, they are based on users’ preferences for items, which can carry a more general meaning than is contained in an item description. Indeed, viewers generally select a film to watch based upon more criteria other than only its genre, director or actors.

Collaborative filtering (CF) techniques depend only on the user's past behavior and provide personalized recommendation of items to users~\cite{linden2003amazon}. CF techniques take a rating matrix as input where the rows of the matrix correspond to users, the columns correspond to items and the cells correspond to the rating given by the user to the item~\cite{RS_ullman_2014}. The rating by a user to an item represents the degree of preference of the user towards the item. The major advantage of the CF methods is they are domain independent. CF methods are classified into Model-based CF~\cite{RS_Koren_2009} and Memory-based CF. Model based techniques model ratings by describing both items and users on some factors induced by matrix factorization strategies. Data sparsity is a serious problem of CF methods. 

Model based approaches such as dimensionality reduction techniques, Principal Component Analysis (PCA), Singular Value Decomposition (SVD) are some of the popular techniques that address the problem of sparsity. However, when certain users or items are ignored, useful information for recommendations may be lost causing reduction in the quality of  recommendation~\cite{CF_Limitation}. The limitation of collaborative filtering systems is the cold start problem. i.e., these methods can not recommend new items to the existing users as there is no past buying history for the item. Similarly, it is difficult to recommend items to a new user without knowing the user's interests in the form of ratings.  

Memory based methods compute a set of nearest neighbors for users as well as items using similarity measures like Pearson coefficient, cosine distance and Manhattan distance. Memory based CF techniques are further classified into user based and item based~\cite{RS_2017}. 

\paragraph{User-based CF}\cite{RS_Taxonomy} User based methods compute the similarity between the users based on their ratings on items~\cite{huang2004graph}. This creates a notion of user neighborhoods. These methods associate a set of nearest neighbors with each user and then predict the user's rating for unrated items utilizing the information of the neighbor's rating for that item.

\paragraph{Item-based CF}~\cite{RS_Item_based}: Similarly, item neighborhood depicts the number of users who rate the same items~\cite{sarwar2001item}. The item rating for a given user can then be predicted based upon the ratings given in their user neighborhood and the item neighborhood. These methods associate an item with a set of similar neighbors, and then predict the user's rating for an item using the ratings given by the user for the nearest neighbors of the target item.

\subsubsection{Hybrid approach}
Hybrid approach uses both types of information, collaborative and content based. Content boosted CF algorithm~\cite{RS_Hybrid1}, uses the item profile information to recommend the items to new users.

\subsubsection{Context aware recommendation system}
Context aware recommendation system is able to label each user action with an appropriate context and effectively tailor the system output to the user in that given context~\cite{context_aware_RS}. In~\cite{kefalas2015graph}, the authors proposed a recommendation of an activity to a social network user based on demographic information. Kefalas et al. have represented their data as a k-partite network modeled as a tensor and proposed novel algorithms. In the same way, in~\cite{ali2020graph}, a citation recommendation is made to researchers also considering tags and time. Standard recommender system algorithms cannot address these issues.

\subsubsection{Graph based Recommender Systems}
Li et al. map transactions to a bipartite user–item interaction graph and thus converting recommendation into a link prediction problem. They propose a kernel-based recommendation approach which generate random walk between user–item pair and define similarities between user–item pairs based on the random walks.  Li et al. use a kernel-based  approach that indirectly examines customers and items related to user-item pair to foresee the existence of an edge between them.  The paper uses a set of nodes representing two types of nodes users(U) and items(I) and edges denoting transactions by the  user regarding the items. The graph kernel is defined on the user-item pairs context~\cite{li2013recommendation}.

Zhang et al. propose a model for music recommendation~\cite{zhang2020bipartite}. The authors represent the recommendation data as a bipartite graph with user and item nodes and the weight of the link is represented as a complex number depicting the dual preference of users in the form of like and dislike and improve the similarity of the users. 

In graph-based recommender system, the task of recommendation is equivalent to predicting a link between user-item based on graphical analysis. Link prediction in graphs is commonly a classification task, that predicts existence of a link in future \cite{LibenNowellBasicLP2007}. But the problem of recommender systems is modeled as a regression task, which predicts the rating of a link. The predicted links with high ratings are generally recommended to users. The standard techniques of recommendation systems take the ratings matrix as input and predicts the empty cells in the matrix. This approach severely suffer with the sparsity in the matrix. Ranking-oriented collaborative filtering approach is more meaningful compared to the rating based approach because, the recommendation is a ranking task recommending top-k ranked items to a user ~\cite{RS_TopN_2010}. In some applications like recommending web pages, rating information may not be available. Graph-based recommendation can efficiently utilize the heterogeneous information available in the networks by expanding the neighborhoods and can compute the proximity between the users and items~\cite{RS_2017}.

A probabilistic measure for predicting future links in bipartite networks is given in~\cite{MyPAKDD}. Several link prediction measures in various kinds of networks have been summarized in~\cite{ASONAM2018}. 

Our aim is to evaluate the efficacy of these measures in the context of recommender systems by modeling them as bipartite graphs. In this paper, we have applied various existing link prediction measures on bipartite graphs in order to build a recommender system. However, the link prediction approach does not give the actual rating, but gives the top k-item recommendations for a user.  For experimentation, we have chosen the bench-mark 'Movielens' dataset. Next section gives an overview of application of link prediction measures on bipartite graphs.

\section{Link Prediction in bipartite graph}

Graph-based recommendation algorithms compute recommendations on a bipartite graph where nodes of the graph  represent users and items and an edge forms between a user node and an item node when the user buys the item. Once the transactions are modeled as a graph, all the standard link prediction methods defined in \cite{LibenNowellBasicLP2007},\cite{LichtenwalterPropFlow2010} can be applied directly on the graph to predict the future links~\cite{RS_JC_2014}. All the measures used in \cite{LibenNowellBasicLP2007} are based on neighbors and paths. Especially, common neighbors, which lie on paths of length two play an important role. All the measures are extended to heterogeneous environment in \cite{ASONAM2018} with a mention of bipartite networks as special case.  

We follow similar notation as in \cite{ASONAM2018} in this paper, which is given below.
\begin{itemize}
\item ($u$, $p$) is a pair of user and product nodes without an edge between them. 
\item $k$  represents hop-distance between two nodes and $t$  denotes time.
\item $\Gamma_k(x)$ is the set of $k$-$hop$ neighbors of node $x$. $\Gamma_1(x)$ refers to the set of all nodes connected by an edge of any type to $x$ generally written as $\Gamma(x)$.  
\item $\Gamma(u)\cap\Gamma(p)$ denotes the set of common neighbors between node $u$ and $p$. In bipartite graph, this set is empty. 
\item $\Gamma_k(u)\cap\Gamma_k(p)$ contains all the common neighbors within $k$-hop distance between nodes $u$ and $p$. In bipartite graph, $k$ needs to be even to start and end path between same type of nodes and odd if path starts and ends between nodes of different types. As recommender system tries to recommend items to users, the odd length paths are meaningful.
\item $P_k(u,p)$ denotes the set of paths connecting $u$ and $p$ by at most $k$ edges.
\end{itemize}

\subsection{Baseline link prediction measures in bipartite graph}
In recommendation systems modeled as bipartite graph, the task is to recommend products to users. The product nodes and user nodes are connected with odd length paths. Even length paths connect the nodes of same type, which is not meaningful in this scenario. Keeping this in mind, the baseline link prediction measures for bipartite graphs are given below. We use suffix B with all link prediction measures to indicate they are used in context of bipartite network.

\begin{itemize}
\item \textbf{Common Neighbors~(CN)} : 
The common neighbor measure in bipartite graph is given as follows:
\begin{equation} \label{CN}
	\begin{split}
		CN_B(u, p)&=|\Gamma_3(u) \cap \Gamma_3(p)| 
	\end{split}
\end{equation}

Jaccard Coefficient, AdamicAdar and Preferential Attachment measures are also neighborhood-based defined in a similar way as follows.
\item \textbf{Jaccard Coefficient~(JC)} : Jaccard Coefficient is the normalized $CN$ measure 
\begin{equation} \label{JC}
	\begin{split}
		JC_B(u, p)= \frac{|\Gamma_3(u) \cap \Gamma_3(p)|}{|\Gamma_3(u) \cup \Gamma_3(p)|} 
	\end{split}
\end{equation}      

\item \textbf{Adamic Adar~(AA)} :  This measure gives importance to the common neighbors with low degree. The following definition for bipartite networks has been hinted at \cite{DarcyMRLP2013}.
\begin{equation} \label{AA}
	\begin{split}
		AA_B(u, p)=\displaystyle \sum_{z \in \Gamma_3(u) \cap \Gamma_3(p)} \frac{1}{log(|\Gamma_3(z)|)} 
	\end{split}
\end{equation}

\item \textbf{Preferential Attachment~(PA)} : This measure does not change in bipartite environment because the measure is concerned only about the degree of the node whatever the type of the node may be. 
\begin{equation} \label{PA}
	PA_B(u, p)=|\Gamma_1(u)| * |\Gamma_1(p)|
\end{equation} 
\item \textbf{Katz~(KZ)}	: This measure is based on the total number of paths between $u$ and $p$ bounded by a limit penalized by path length. 

\begin{equation} \label{KZ}
	\begin{split}
		KZ_B(u, p)= \sum_{l} \beta ^l |P_l{(u, p)}^{l}| 
	\end{split}	
\end{equation}	

\item \textbf{Page Rank~(PR)} : This measure can be extended to heterogeneous network by including the heterogeneous edges in the random-walk.
\begin{equation}\label{PR}
	\begin{split}
		PR_B(u)=\frac{1-\alpha}{|E|}+\alpha\displaystyle \sum_{z \in \Gamma_3(u)}\frac{PR(z)}{|\Gamma_3(z)|}
	\end{split}
\end{equation}
where $|E|$ is the total number of links in $G$.

\item \textbf{Rooted Page Rank~(RPR)} : In order to make the measure $PR$ symmetric, Rooted Page Rank is computed for an edge $(u, p)$ as follows. 

\begin{equation}
	RPR_B(u, p)=PR(u, p)+PR(p, u)
\end{equation}

\item \textbf{PropFlow~(PF)}	: This is a random-walk beginning  at node $u$ and ending at $p$ within $l$ steps. This random-walk from node $u$ to $p$ terminates either when it reaches $p$ or revisits any node. $PF_B(u, p)$ is the probability of information flow from $u$ to $p$ based on random transmission along all paths defined recursively as follows.

\begin{equation} \label{PF}
	PF_B(u, p) = \displaystyle\sum_{l=2}^L \displaystyle\sum_{p \in paths_l(u, p)} \sum_{\forall (z_1,z_2) \in p} PF(z_1,z_2) 
\end{equation}

where 

\begin{equation} \label{PF_rec}
	PF_B(z_1,z_2) = PF(a,z_1) * \frac{w(z_1,z_2)}{\displaystyle\sum_{z \in \Gamma(z_1)}w(z_1,z)} 
\end{equation}
with $a$ as previous node of $z_1$ in the random-walk, $PF(a,z_1)$=1 if $a$ is the starting node and $paths_l(u, p)$ is the set of paths of length $l$ between $u$ and $p$. In bipartite network, $l$ is odd. 

\begin{equation} \label{HPF}
	PF_B(u, p) = \displaystyle\sum_{l=2}^L \displaystyle\sum_{p \in P_l(u, p)} \sum_{\forall(z_1,z_2) \in p} PF_B(z_1,z_2) 
\end{equation}

where $PF_B(z_1,z_2) $ is as defined in Eq.\ref{PF_rec}. 
\end{itemize}

\subsection{Temporal measures for link prediction in bipartite graph}

\subsubsection{Time-Score($TS_B$)}

We extend Time-Score measure proposed for homogeneous networks~\cite{TimeScore} to bipartite environment as follows: 
\begin{equation}
	\label{Hetero-TS}
	TS_B(u,p)=  \sum_{path \in P_3(u,p)}\frac{w(path)* \beta^{r(path)}}{|latest(path)-oldest(path)|+1} 
\end{equation}

where  $w(path)$ is equal to the harmonic mean of edge weights of edges in $path$, $\beta$ is a damping factor~(0<$\beta$<1),  
$latest($path$)=\displaystyle \max_{e~on~P_3}(t(e))$, $oldest($path$)= \displaystyle \min_{e~on~P_3}(t(e))$ and $r$ is a recency factor, defined as~~ $r($path$)$= $current\_time-latest($path$)$.

\subsubsection{Link-Score($LS_B$)} 

Choudhary et al. extend the Time-score measure to obtain a path based measure called $Link$-$score$~\cite{LinkScore} for homogeneous networks. To obtain the $Link$-$score$ between a pair of nodes $u$ and $p$ which are not directly connected, the authors define a Time Path Index~($TPI$) on each path $p$ between the nodes $u$ and $p$. $TPI$ evaluates path weight based on time stamps of links involved in a path. Link-score is the sum of $TPI$ of each path between the nodes $x$ and $y$.  

\hspace{3em}We extend $Link$-$score$ to bipartite network by considering $paths$ of odd length between two nodes instead of paths containing any length. The modified definitions of $TPI$ and $Link$-$Score$  are given in equation~\ref{Hetero-LS}.

\begin{equation}
	\label{Hetero-LS}
	TPI_{path}=  \frac{w(path)* \beta^{current\_time-avg(path)}} {|current\_time-latest(path)|+1} 
\end{equation}

where $avg(path)$ is the average active year, which is the average of years of recent interaction of edges on odd length path and all others are as defined in equation~\ref{Hetero-TS}.

\begin{equation}
	LS_B(u,p)=\displaystyle \sum_{l=2}^{L}\frac{Avg(TPI_{P_l(u,p)})}{l-1}
\end{equation}

where $L$ is the maximum length of paths between nodes $i$ and $j$.

\subsubsection{T\_Flow($TF_B$)}

T\_Flow~\cite{TFlow} is a random-walk based measure, which is an extension of PropFlow measure defined in~\cite{LichtenwalterPropFlow2010}. Munasinghe et.al~\cite{TFlow} define T\_Flow that computes the information flow between a pair of nodes $x$ and $y$ through all random-walks starting from node $u$ to node $p$   including link weights as well as activeness of links by giving more weight to recently formed links recursively and take the summation.

We extend T\_Flow measure to bipartite networks as follows: 

\begin{equation}
	\label{Hetero-TF}
	TF_B(u,p)=\displaystyle\sum_{i=2}^l \displaystyle\sum_{path \in P_l(x,y)} \sum_{\forall e \in path} TF_B(z_1,z_2) * (1-\alpha)^{r(path)}
\end{equation}
If $(u,p)\in E$, then $TF_B(u,p)$ is given by 
\begin{equation}
	TF_B(u,p) = TF_B(a,u) * \frac{w(u,path)}{\displaystyle\sum_{z \in \Gamma(u)}w(u,z)}*(1-\alpha)^{r(path)} 
\end{equation}
where $t_u$ is the time stamp of the link when the random walk visits the node $u$ and $t_p$ is the time stamp of the link when the random walk visits node $p$.

\subsection{Probabilistic measure for link prediction in bipartite graph}
The Probabilistic Graphical Model (PGM) represents the structure of the graph in a natural way by considering the nodes of graph as random variables and edges as dependencies between them. By representing a graph as a PGM, the problem of link prediction, which is calculation of the probability of link formation between two nodes $u,~p$ is translated to computing the joint probability of the random variables $U,~P$.

Recommender systems represented as bipartite networks are large in size. Therefore, finding the joint probabilities of link formation is intractable. But the links in the graphs are also sparse, with nodes generally directly connected to  only a few other nodes. For example, in the e-commerce sites, out of lakhs of items, a user buys only a few items. This property allows the PGM distribution to be represented tractably. The model of this framework is simple to understand. Inference between two random variables is same as finding the joint probability between those two random variables in PGM. Many algorithms are available for computation of joint probability between variables, given evidence on others. These inference algorithms work directly on the graph structure and are generally faster than computing the joint distribution explicitly. With all these advantages, link prediction in PGM is more effective.

In most of the cases, the pair-wise interactions of entities are available in the event logs. For example, a user buying/rating three items say $p_1$, $p_2$ and $p_3$ is available in the corresponding transaction database. In order to model the unknown distribution of co-occurrence probabilities, events available in the event logs can be used as constraints to build a model for the unknown distribution. Probabilistic Graphical Models efficiently utilize this higher order topological information and thus are efficient in link prediction task~\cite{Cooccur:ICDM:2007}. 

The probabilistic model helps in estimating the joint probability distribution over the variables of the model~\cite{JointPDF}. That means, a probabilistic model represents the probability distribution over all possible  deterministic states of the model. This facilitates the inference of marginal probabilities of the individual variables of the model.

Wang et al. \cite{Cooccur:ICDM:2007} are among first researchers who modeled the problem of link prediction using MRFs. Kashima et al. \cite{Kashima} propose a probabilistic model of network evolution and apply it for predicting future links. The authors show that by intelligently selecting the parameters in an evolution model, the problem of network evolution reduces to the problem of link prediction.  Clauset et al. \cite{Clauset_2008_Hierarchical} propose a probabilistic model based on hierarchical structure of the network. In hierarchical structure, the vertices are divided into groups that further subdivide into groups of groups and so on. The model infers hierarchical structure from network data and can be used for prediction of missing links. The learning task uses the observed network data and infers the  most likely hierarchical structure through statistical inference.

The works of Kashima~\cite{Kashima} and Clauset~\cite{Clauset_2008_Hierarchical} are global models and are not scalable for large networks. The method of Wang et. al. \cite{Cooccur:ICDM:2007} uses local probabilistic information of graphs for link prediction and we adopt their method in our work. The following section explains the algorithm for link prediction using MRF proposed by Wang et al. \cite{Cooccur:ICDM:2007}

\section{Proposed approach: Link Prediction in bipartite networks induced as PGM}
Wang et al. \cite{Cooccur:ICDM:2007} propose a measure called Co-occurrence Probability (COP) to be computed between a pair of nodes without an edge between them. The procedure for computing COP has three steps. First step chooses a few set of nodes which may play a significant role in future link formation. Second step computes a Markov Random Field with the chosen nodes in the first step. Third step infers joint probability between the two nodes using the MRF constructed. The measure has been extended to temporal networks in ~\cite{TCOP}. Heterogeneous version of the measure is defined in ~\cite{MyPAKDD}. Though $Hetero$-COP has been defined in~\cite{MyPAKDD}, the algorithm has to be re-worked for the bipartite context. In this paper, we have extended COP measure to bipartite environment, named it as $B$-COP of a missing link $(u, p)$ in the lines of $COP$, which is given in Algorithm \ref{Hetero-COP-Algo}. The extended computation procedure for bipartite networks is given in subsequent sections.

\begin{algorithm}[H]
	\caption{$B$-COP measure for Link Prediction in bipartite graphs}
	\label{Hetero-COP-Algo}
	\textbf{Input}: $G = (V, E),$ where \\ 
	\hspace{\algorithmicindent} $V=\lbrace V_1 \cup V_2 \rbrace$ is the set of two types of nodes,  \\ 
	\hspace{\algorithmicindent} $E$ is the set of links from nodes of $V_1$ to $V_2$. \\
	\textbf{Output}: $B$-COP$(u, p)$ where $(u, p)$ is a missing link.\\ 
	
	\begin{algorithmic}
		\STATE {\bf Step 1:}	 Extract $B$-cliques from $G$, from the event logs using Algorithm \ref{LP-based-RS}. Call it $BCliq$.
		\STATE {\bf Step 2:}	 Compute \textbf{BCNS}, the central neighborhood set of $(u, p)$ using the Algorithm \ref{HCNS-Computation}. 
		\STATE {\bf Step 3:}	 Extract $B$-cliques formed with the nodes in BCNS and compute clique potentials. This forms the local MRF with the nodes in BCNS.
		\STATE {\bf Step 4:}     Return $B$-COP$(u, p)$ which is the joint probability of link $(u, p)$ using junction tree algorithm.	
	\end{algorithmic}
\end{algorithm}

\begin{figure}[h]
	\centering
	\includegraphics[width=0.25\textwidth]{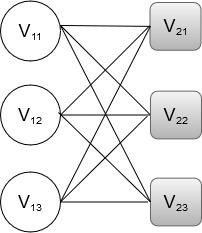}
	\caption{An example $B$-clique.}
	\label{B-clique}
\end{figure}

\hspace{3em}A clique in bipartite graphs can be considered as complete bipartite sub-graph. A sample $B$-clique is shown in Fig.\ref{B-clique}.

\subsection{Computation of Central Neighborhood Set in Bipartite graph(BCNS)}
BCNS$(u,p)$ is computed using a breadth first search based algorithm on bipartite graph as follows: All paths between $u$ and $p$ are obtained using breadth first search (BFS) algorithm. Then, the frequency score of each path is computed by summing the occurrence count of nodes on the path. Occurrence count of a node is the number of times the node appears in all paths. The paths are now ordered in the increasing order of length with equal paths being ordered in decreasing order of frequency score.  The size of the central neighborhood set is further restricted by considering only top $k$ nodes. The procedure of computing $BCNS(u,p)$ is described in the Algorithm~\ref{HCNS-Computation}.

\begin{algorithm}[!htb]
	\caption{ {\bf Bipartite Central Neighborhood Set($G, u, p, l,maxSize$)}}
	\label{HCNS-Computation}
	
	\textbf{Input}: $G$: a graph; $u$: starting node; $p$: ending node; $l$:maximum path length; $maxSize$:Central Neighborhood Set size threshold \\
	\textbf{Output}: $BCNS$, Bipartite Central Neighborhood Set between $u$ and $p$; \\ 
	
	\begin{algorithmic} 
		\STATE  {\bf Step 1:} \hspace{\algorithmicindent} Compute paths of length $\leq~ l$  between $u$ and $p$.
		\STATE  {\bf Step 2:} \hspace{\algorithmicindent} Find occurrence count $O_k$ of each node $k$ in paths between $u$ and $p$.
		\STATE  {\bf Step 3:}\hspace{\algorithmicindent} Compute $frequency$-$score$ $F_p$, of each path $p$ as follows: \\
		\begin{center}
			$F_p=\sum_{k \in p} O_k$ \\
		\end{center}
		\hspace{\algorithmicindent} $frequency$-$score$ of a path is the sum of the occurrence counts of all nodes along the paths. 
		\STATE {\bf Step 4:} \hspace{\algorithmicindent} Sort the paths in increasing order of path length and then in decreasing order of  $frequency$-$score$. Let the ranked list of paths be $P$. 
		\STATE {\bf Step 5:} \hspace{\algorithmicindent}
		
		\WHILE {size(BCNS) $\leq maxSize$} 
		\STATE Add nodes of path $p \in P$ to BCNS
		\ENDWHILE
		\RETURN BCNS
	\end{algorithmic}
\end{algorithm} 
 
The procedure of computation of BCNS for a toy example given  in Fig.\ref{CNS-toy} is shown in Table~\ref{CNS-paths}.

\begin{minipage}{\textwidth}
	\begin{minipage}[c]{0.39\textwidth}
		\includegraphics[width=0.7\textwidth]{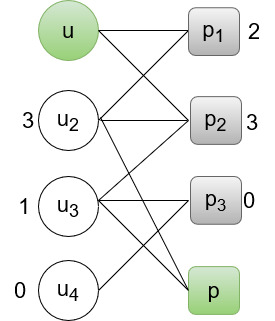}     
		\captionof{figure}{\small{A toy example for illustrating computation of BCNS between nodes $u$ and $p$. Node weights are the occurrence counts.}}
		\label{CNS-toy}
	\end{minipage}
	\hfill
	\begin{minipage}[c]{0.59\textwidth}
		\small{
			\begin{tabular}{|l|l|} \hline
				\textbf{path(p)}   &  \textbf{frequency-score($F_p$)} \\ \hline
				$u-p_2-u_2-p$     & $O_{p_2} + O_{u_2}$ = 6             \\
				$u-p_1-u_2-p$     & $O_{p_1} + O_{u_2}$ = 5             \\
				$u-p_2-u_3-p$     & $O_{p_2} + O_{u_3}$ = 4             \\
				$u-p_1-u_2-p_2-u_3-p$     & $O_{p_1} + O_{u_2} + O_{p_2} + O_{u_3}$ = 9     \\ \hline
			\end{tabular}
		}
		\captionof{table}{All paths between nodes $u$ and $p$ in Fig.\ref{CNS-toy} and their frequency scores.}\label{CNS-paths}
	\end{minipage}
\end{minipage}

Table~\ref{CNS-paths} shows all paths sorted in the increasing order of path length and frequency score, between the nodes $u$ and $p$ of the graph shown in Fig.\ref{CNS-toy} along with their $frequency$-$scores$.
 One can observe that all these paths are of odd length. For the bipartite graph in Fig.\ref{CNS-toy}, BCNS$\left(u,p\right)= \lbrace u,p_2,u_2,p_1,p \rbrace$, if $maxSize$ is taken as 5. 
\subsection{Construction of local MRF} 
After computing the BCNS, the $B$-cliques containing only nodes of  BCNS are extracted. This forms the clique graph of local MRF. MRF construction needs computation of clique potentials. The clique potential table of a $B$-clique is computed using the NDI.

In most of the cases, the information of homogeneous cliques containing all nodes of same type are available in the event logs. For example, in coauthorship networks, the group of authors who publish a paper together forms a homogeneous clique of author nodes and an author who publishes a paper in a conference forms a heterogeneous edge between the author node and conference node. But in the case of recommender systems, user cliques and item cliques are not readily available in the event logs. We propose an algorithm for extracting B-cliques shown in Algorithm~\ref{LP-based-RS}. Since the number of users is huge in comparison to the set of items which is a much smaller set, the extraction of $B$-cliques can start from the item cliques. 

\begin{algorithm}
	\caption{\bf Extraction of \textbf{B-Cliques} from user-item bipartite graph} \label{LP-based-RS}
	
	\textbf{Input}:  $G=(V,E)$ where $V = U \cup I$, $U$ is set of user nodes and $I$ is the set of item nodes and $E \subseteq U X I$ represents weighted edges. \\
	\textbf{Output}: $Bcliq$, set of maximal B-cliques of $G$. 

	\begin{algorithmic}
		
		\STATE $\textbf{Step 1:}$ Extract the set of all item cliques, $Item\_Cliq$ as follows:
		\STATE $Item\_Cliq=\phi$ 
		\FOR {each user $u \in U$} 
		\STATE $I_u=\phi $
		\FOR {each $i \in I \text{~~and~~} (u,i)\in E$} 
		\STATE $I_u=I_u \cup \{ i \}$ 
		\ENDFOR ~~//~~$I_u$ is formed by taking all the items to which $u$ gives a rating. \\
		$Item\_Cliq=Item\_Cliq \cup I_u$
		\ENDFOR
		
		\STATE $\textbf{Step 2:}$ Extract the set of all user cliques, $User\_Cliq$ as follows:
		\STATE $User\_Cliq=\phi$ 
		\FOR {each item $i \in $} 
		\STATE $U_i=\phi $
		\FOR {each $u \in U \text{~~and~~} (u,i)\in E$} 
		\STATE $U_i=U_i \cup \{ u \}$ 
		\ENDFOR ~~//~~$U_i$ contains all users that have rated item $i$. \\
		$User\_Cliq=User\_Cliq \cup U_i$
		\ENDFOR
		
		\STATE $\textbf{Step 3:}$  
		\FOR {each item $i$ in $I$ }
		\STATE $J \leftarrow I$ 
		\FOR {each user $v$ in $U_i$ }
		   \STATE $J \leftarrow  J \cap ~I_v$
		\ENDFOR
		\STATE	//~~ $J = \displaystyle\bigcap_{v \in U_i} I_v$ ~~// 
		\STATE 	$Bcliq_i = J \cup U_i$
		\STATE $Bcliq  = \cup_i ~ Bcliq_i$
		\ENDFOR
		\STATE return $Bcliq$
	\\	
	\end{algorithmic}
\end{algorithm}

\begin{figure}[H]
	\centering
	\includegraphics[width=1.2\textwidth]{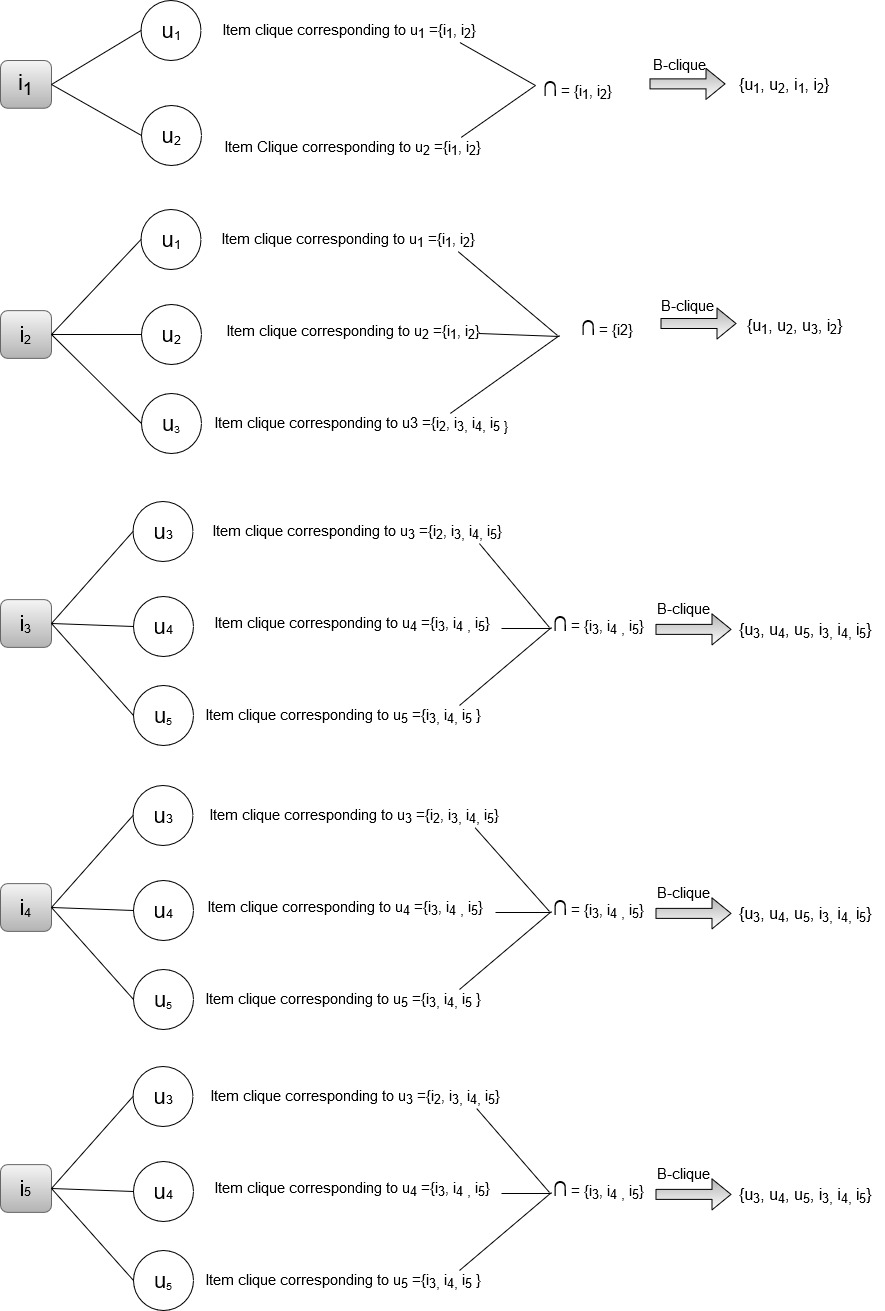}
	\caption{Illustration of extracting $B$-cliques from User-Item event logs}
	\label{extract-item-b-clique}
\end{figure}

The B-clique extraction algorithm first extracts all homogeneous cliques of items $Icliq$ and users $Ucliq$. A homogeneous user clique is formed with all the users who rate/buy the same item. Similarly, a homogeneous clique of items is formed with all the items a user rate/buy. To extract a B-clique, first consider an item $i$. For each user $v$ who have rated $i$, compute the common items rated by user $v$. The union of $U_i$ along with all the common items rated by $U_i$ forms a $B$-clique. The process of extracting B-cliques for toy example in Fig.\ref{user-item-bipartite} is illustrated below:
\vspace{5em}

\begin{minipage}{\textwidth}
	\begin{minipage}[l]{0.3\textwidth}
		\begin{figure}[H]
			\includegraphics[scale=0.55]{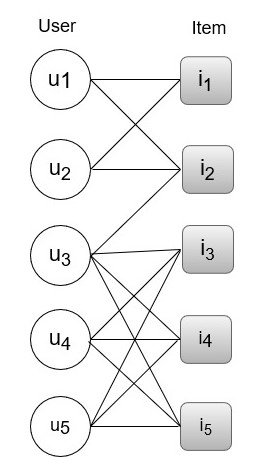}
			\caption{User-Item bipartite graph}
			\label{user-item-bipartite}
		\end{figure}
	\end{minipage}
	\hfill
	\begin{minipage}[c]{0.68\textwidth}
		U=$\lbrace u_1,u_2,u_3,u_4,u_5 \rbrace$ \hspace{2cm} I=$\lbrace i_1,i_2,i_3,i_4,i_5 \rbrace$ \\
		Item cliques : \\
		Item clique corresponding to user $u_1=\lbrace i_1, i_2 \rbrace$ \\
		Item clique corresponding to user $u_2=\lbrace i_1, i_2 \rbrace$ \\
		Item clique corresponding to user $u_3=\lbrace i_2, i_3, i_4, i_5 \rbrace$ \\
		Item clique corresponding to user $u_4=\lbrace 1_3, i_4, i_5\rbrace$ \\
		Item clique corresponding to user $u_5=\lbrace 1_3, i_4, i_5\rbrace$ \\
		User cliques : \\
		User clique corresponding to item $i_1=\lbrace u_1, u_2 \rbrace$ \\
		User clique corresponding to item $i_2=\lbrace u_1, u_2, u_3 \rbrace$ \\
		User clique corresponding to item $i_3=\lbrace u_3, u_4, u_5 \rbrace$ \\
		User clique corresponding to item $i_4=\lbrace u_3, u_4, u_5 \rbrace$ \\
		User clique corresponding to item $i_5=\lbrace u_3, u_4, u_5 \rbrace$ \\ 
		The the extraction of B-cliques from the above homogeneous cliques is shown in Fig.\ref{extract-item-b-clique}. 
	\end{minipage}
\end{minipage}

\subsection{Computation of $B$-COP score} Once the local MRF of a pair of nodes $u$ and $p$  is constructed, the $B$-COP score between the nodes is obtained using junction tree inference algorithm. Note that $B$-COP score for a link $u$ - $p$ cannot be computed if $u$  and $p$ are in disjoint cliques as there exists no path connecting these cliques.  

The experimental evaluation of proposed approach is given in next section.

\section{Experimental Evaluation}
The experimentation is carried out on a benchmark MovieLens recommender system whose details are given below.

\subsection{Dataset}
This data set used for experimental evaluation contains more than ten million ratings given by 71,567 users on 10,681 movies of the online movie recommender service MovieLens~\cite{movielens}. In the graph recommendation systems, movies are considered as nodes and items are users and rating given by users to the movies are considered as the weights on links. In MovieLens dataset, the train-test splits are given in \cite{movielens}. The users who have rated at least 20 movies have been chosen randomly to be included in this dataset. The benchmark data set~\cite{movielens1} is given with 80\% - 20\% split with 80\% given as training set and test set containing 20\%. The training and test sets are formed by splitting the ratings data such that, for every user, 80\% of his/her ratings is taken in training and the rest are taken in the test set. In this experimentation, 5 fold cross validation is used. All the 5 sets of training and test datasets are made available at \cite{movielens1}.
The evaluation metrics used for recommender systems are given in the following section.

\subsection{Evaluation Metrics}
Evaluation metrics in recommender systems can be classified as 
\begin{itemize}
	\item Accuracy measures: Mean Absolute Error (MAE), Root of Mean Square Error (RMSE), Normalized Mean Average Error (NMAE).
	\item Set recommendation metrics : Precision, Recall and Area Under Receiver Operating Characteristic (AUROC), Area Under Precision-recall curve (AUPR)
	\item Rank recommendation metrics: Half-life, discounted cumulative gain and Rank-score
\end{itemize}

\hspace{3em}Most of the measures listed above, use rating to calculate the error and hence are not applicable in our context. We use AUROC, AUPR are used for evaluating performance.

\paragraph{Rank-score} Rank-score metric measures the ability of a recommendation algorithm to produce a ranked list of recommended items. The recommender system method is efficient, if the ranking given by the method matches with the user's order of buying the items in the recommended list. Rank-score is defined as follows : For a user $u$, the list of items $i$ recommended to $u$, that is predicted by the algorithm is captured by $rankscore_p$
\begin{center}
\begin{math}
	\begin{aligned}
		rankscore_{max}   = \frac{1}{\sum_{j =1}^{|T|} 2^{\frac{j-1}{\alpha}}} \\
		rankscore_p       = \frac{1}{\sum_{j \in T} 2^{\frac{rank(j)-1}{\alpha}}} \\
		Rankscore         = \frac{rankscore_p}{rankscore_{max}} \\  
	\end{aligned}
\end{math}
\end{center}

where $rank(j)$ is the rank given by the recommender algorithm to item $j$. $T$ is the set of items of interest and $\alpha$ is ranking half-life, an exponential reduction factor and can be any numeric. 

\subsection{Results and Discussion}
The prediction scores of all baseline link prediction measures are computed using the tool $LPMade$ \cite{lpmade}, with default parameters.
In the computation of $TS_B$, $LS_B$ and $TPI$, the damping factor $\beta$ is taken as 0.5. The decay factor $\alpha$ is considered as 0.1 in computation of $TF_B$. In $TCOP$, maximum path length of 10 is taken for computing BCNS. The accuracy of all these link prediction measures is compared with the standard User-based and Item-based collaborative filtering (CF) methods. Pearson correlation coefficient is used to find the similar users in User-based CF and cosine similarity is used for finding item similarity in Item-based CF.

The results obtained for AUROC, AUPR and Rank-score for MovieLens dataset are given in Table.\ref{movielens-results}. 

\begin{table}[h]
	\centering
	\caption{Performance of LP measures for recommending movies to users in \textbf{MovieLens} Bipartite Network}
	\label{movielens-results}
	\begin{tabular}{|l||l|l|l|}
		
		\hline
		\textbf{LP measure} & \textbf{Rank-score} & \textbf{AUROC} & \textbf{AUPR} \\ \hline \hline
		\multicolumn{4}{|c|}{\textbf{Non-temporal} LP measures} \\ \hline \hline
		$CN_B$               & 2.0501              & 0.5123         & 0.0092        \\ \hline
		$JC_B$               & 1.3615              & 0.4956         & 0.0085        \\ \hline
		$AA_B$               & 1.7819              & 0.5749         & 0.0153        \\ \hline
		$PA_B$               & 2.6804              & 0.6832         & 0.0251        \\ \hline
		$KZ_B$               & 3.2546              & 0.6635         & 0.0193        \\ \hline
		$PF_B$               & 3.2990              & 0.6846         & 0.0286        \\ \hline
		$COP_B$              & 3.6661              & 0.7231         & 0.0613        \\ \hline \hline
		\multicolumn{4}{|c|}{\textbf{Temporal} LP measures} \\ \hline \hline
		$TS_B$               & 4.0015              & 0.7016         & 0.0365        \\ \hline
		$LS_B$               & 5.2134              & 0.7340         & 0.0861        \\ \hline
		$TF_B$               & 5.3684              & 0.7532         & 0.0960        \\ \hline
		$TCOP_B$             & {\color{red}\textbf{8.6304}}     & {\color{red}\textbf{0.8165}}       & {\color{red}\textbf{0.2351}}        \\ \hline \hline
\multicolumn{4}{|c|}{\textbf{Classical Collaborative Filtering} methods} \\ \hline \hline
User-based CF    & 3.9942              & 0.6925         & 0.0415        \\ \hline
Item-based CF    & 3.0274              & 0.7136         & 0.0491        \\ \hline
\end{tabular}
\end{table}

\begin{figure}
\centering
\includegraphics[width=1\textwidth]{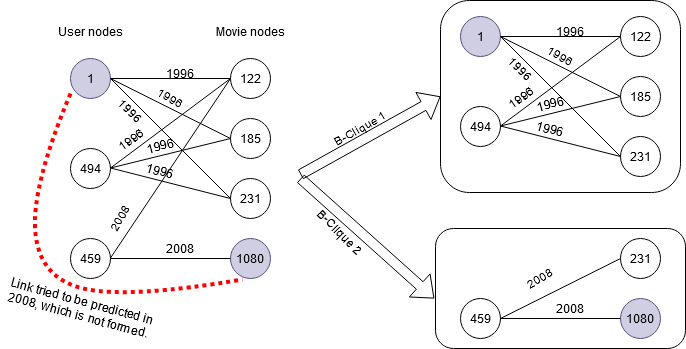}
\caption{A snapshot from movielens dataset where the link predicted is not formed.}
\label{Movie_neg}
\end{figure}

\begin{figure}
\centering
\includegraphics[width=1\textwidth]{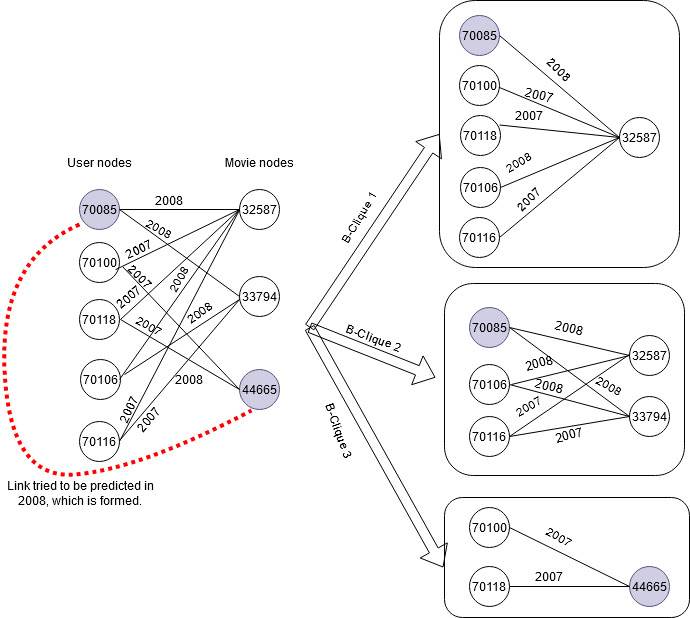}
\caption{A snapshot from movielens dataset where the link predicted is formed.}
\label{Movie_pos}
\end{figure}

First observation in this experimentation is that some of the link prediction measures like Katz, PropFlow could produce better recommendation compared to Item-based CF. The usage of temporal measures seem to help in improving the quality of recommendations. The time of formation of link or the time of rating given by a user to movie plays crucial role, as the user's preferences change over time. The temporal measures TS, LS, TF and TCOP assign more weight to the recent ratings. Therefore, temporal measures performed better than all the other measures including User-based CF and Item-based CF. Fig.\ref{Movie_neg} depicts a situation where non-temporal measures predict a link between the nodes $1$ and $1080$ and temporal measures predict correctly that the link won't be formed. This is clearly evident from the fact that $B-Cliq 1$ is formed with very old links formed in the year 1996 compared to the prediction year 2008. Fig.\ref{Movie_pos} depicts the situation where the temporal links predict correctly. TCOP outperformed all the other link prediction measures and the user-based and item-based collaborative filtering methods. The recommendation performance is improved by 6\% in terms of AUROC over TF and nearly 10\% over CF methods. It can be observed from Table.\ref{movielens-results} that the AUPR score also have shown great improvement from 0.0960(of TF) to 0.2351. Similar trend is observed for the  evaluation measure of  Rank-score. All the temporal measures perform better than User-based CF and Item-based CF. TCOP is rated as highest by Rank-score.

\section{Conclusion}
In this paper, the recommender systems problem is solved using link prediction approach. Link prediction approach is scalable as it is based on local neighborhood of the large sparse graph. The standard recommender systems approaches do not utilize the temporal information available on the link effectively. In this work, some extensions are proposed to existing temporal measures in bipartite graphs. One of the main contributions of this work is an algorithm for computing temporal cooccurrence probability measure on bipartite graphs and its application to movie recommendation system. Temporal measures for link prediction such as Time score, Link Score, T\_Flow and Temporal cooccurrence measure achieve improvement in recommendation quality by utilizing this temporal information more efficiently. However, link prediction approach to solve recommender systems do not address the cold start problem. In future, we would like to work on predicting actual rating of the link.
\bibliographystyle{aps-nameyear} 
\bibliography{LP_approach_to_RS}

\end{document}